\documentclass[12pt]{article}
\usepackage{epsfig}
\def\be{\begin{equation}}
\def\ee{\end{equation}}
\newcommand{\bea}{\begin{eqnarray}}
\newcommand{\eea}{\end{eqnarray}}

\newcommand{\bp}{{\bf p}}

\newcommand{\RSCHNC}{R_{\mbox{\tiny{SCHNC}}}}
\date{\empty}

\def\lsim{\mathrel{\rlap{\lower4pt\hbox{\hskip1pt$\sim$}}
    \raise1pt\hbox{$<$}}}         

\def\gsim{\mathrel{\rlap{\lower4pt\hbox{\hskip1pt$\sim$}}
    \raise1pt\hbox{$>$}}}         

\title{Production of orbitally excited vector mesons in diffractive DIS}

\author{F. Caporale$^{1}$, I.P. Ivanov$^{1,2}$\\
  {\normalsize $^1$ INFN, Gruppo Collegato di Cosenza, Italy}\\
  {\normalsize $^2$ Sobolev Institute of Mathematics,  Novosibirsk, Russia}}

\begin{document}
\maketitle

\begin{abstract}
Within the $k_t$-factorization framework, we study diffractive production 
of orbitally excited vector mesons and compare it with the production 
of radial excitations, focusing on the $\rho(1450)/\rho(1700)$ case.
At small $Q^2$, orbital excitation of light quarkonia
is found to dominate over radial excitations in diffractive production.
We predict strong suppression of the production of orbital excitations 
by longitudinal photons, which leads to very small $\sigma_L/\sigma_T$ ratio.
At small $Q^2$, the $s$-channel helicity violating transitions
contribute $\sim 10$--15 $\%$ of the transverse cross section and $\sim 50\%$
of the longitudinal cross section.
We also study mixing between radial and orbital excitations
and determine strategies towards clarification of $S$-wave/$D$-wave assignment
to $\rho(1450)$ and $\rho(1700)$ mesons.
The results are compared with the experimental data available, and
predictions for future experiments are given. 
\end{abstract}

\section{Introduction}

Diffractive production of vector mesons (VM) in DIS $\gamma^* p \to Vp$ ($V = \rho,\phi,J/\psi$ etc.)
is a very active field of research (see recent review \cite{review} and references therein). 
So far, the main focus has been on the ground state mesons,
while diffractive production of excited states has not enjoyed much attention.
Perhaps, the most studied case so far was the production of radially excited charmonium 
$J/\psi(2S)$, where
remarkable consequences of the presence of a node in the radial wave function
\cite{2s-th}  were nicely confirmed by H1 measurements \cite{psi2s-ex}.

Experimental studies of excited $\rho'$ production should be 
even more rewarding.
First, production of excited mesons probes the dipole cross section 
at larger dipole sizes than the
production of ground states, see e.g. analysis of $\rho'$/$\rho''$ in \cite{dosch}
and the recent study of $\rho_3$ in \cite{spin3}.
Such sensitivity to soft diffraction can help understand the phenomenon of 
saturation,  \cite{saturation}. 
Besides, it is expected that diffractive production of $\rho'$
can help resolve the long standing puzzle of the radial/orbital excitation
assignment to the $\rho(1450)$ and $\rho(1700)$ mesons\footnote{In 
his analysis of $\rho(1450)$ and $\rho(1700)$ mesons, 
the author of \cite{4mesons} concludes that 
"The way forward with $1^{--}$ $q\bar q$ states is to study diffractive 
dissociation of the photon".}, see \cite{4mesons}.

Diffractive production of excited $\rho'$ mesons
has been observed in a number of fixed target experiments 
with relatively high energies.
Diffractive production of $\rho'(1600)$ was reported in $\pi^+\pi^-$ 
\cite{aston1980} and $4\pi$ \cite{atkinson1981} final states
(for reanalysis of these data in terms of $\rho(1450)$ 
and $\rho(1700)$ mesons see \cite{tworhoprimes}). 
These states were also studied in the Fermilab experiment E687 \cite{E687}
both in $2\pi$ and $4\pi$ channels. In addition, there are indications
that the $\rho(1700)$ --- which is believed to be predominantly the orbital excitation --- 
can play essential role in an interesting narrow dip structure
observed by E687 in the $6\pi$ final state \cite{E687-6pi}.
However, all these experiments gave only the value of the photoproduction cross section,
and neither energy or $Q^2$ dependence, nor helicity structure
of the reaction was studied. This gap was partially filled by the 
H1 measurement of $\rho'$ electroproduction at $4 < Q^2 < 50$ GeV$^2$ \cite{H1},
but due to low statistics the results presented had large errorbars.

In theory, production of orbitally excited vector mesons
is expected to be suppressed by Fermi motion, 
as its radial wave function vanishes at the origin. 
This suppression was believed to be sufficiently strong and 
even prompted the authors of \cite{dosch} to consider diffractive $\rho(1450)$ and $\rho(1700)$
production neglecting in both cases the $D$-wave contributions altogether.
Only in \cite{in99} were the $S$-wave and $D$-wave vector meson
production amplitudes calculated within the $k_t$-factorization approach,
and the first estimates \cite{phd} showed that at small-to-moderate $Q^2$
the production rates of the $D$-wave and $2S$ $\rho'$ states were roughly of the same order.

In this paper we extend this research and report more detailed numerical 
results on $D$-wave vector meson production.

\section{Amplitude of the vector meson production}

We use the usual notation for kinematical variables. $Q^2$ is 
the photon's virtuality, $W$ is the total center-of-mass energy of the $\gamma^*p$ collision.
The momentum transfer from proton to photon is denoted by $\Delta_\mu$ and at high energies 
is almost purely transverse: $-\Delta^2 = |t| \approx |t'| = \vec \Delta^2$. 
The transverse vectors (orthogonal to the $\gamma^*p$ collision axis)  
will be always labelled by an arrow.

Diffractive production of meson $V$ with mass $m_V$ can be treated in the lowest
Fock state approximation 
as production of the $q\bar q$ pair of invariant mass $M\not = m_V$, 
which is then projected, at the amplitude level, onto the final state.
Within the leading log${1 \over x}$ accuracy the higher Fock states are reabsorbed into
the evolution of the unintegrated gluon density (or color dipole cross section).
A typical amplitude contains the valence quark loop,
with integration over the quark transverse momentum $\vec k$ 
and its fraction of photon's lightcone momentum $z$,
and the uppermost gluon loop, 
with the integration over transverse momentum $\vec \kappa$. 

Throughout the text, the ground state vector mesons 
(always understood as $1S$ states)
will be generically labelled by $V$ or $V_{1S}$, their radial excitations will be labelled
by $V_{2S}$, while the pure $D$-wave vector mesons will 
be labelled by $V_D$.

A generic form of the helicity amplitudes 
$\gamma^*(\lambda_\gamma) \to V_D(\lambda_V)$ is
\be
Im A_{\lambda_V;\lambda_\gamma} =  {c_V \sqrt{4\pi\alpha_{em}} \over 4 \pi^2} 
\int {dzd^2\vec k\over z(1-z)}  \int {d^2\vec \kappa \over \vec \kappa^4}
\alpha_s\,{\cal F}(x_1,x_2,\vec \kappa,\vec\Delta)\cdot
I(\lambda_\gamma\to \lambda_V)\cdot\psi_D(\bp^2)\,.\label{ampl}
\ee
Here $c_V$ is the flavor-dependent average charge
of the quark, the argument of the strong coupling constant 
$\alpha_s$ is max$[z(1-z)(Q^2+M^2),\vec\kappa^2]$.
The radial wave function of the vector meson depends on the
spherically symmetric quantity $\bp^2$, where $\bp$ is the relative $q\bar q$ 
momentum written in the $q\bar q$ rest frame.  
${\cal F}(x_{1},x_{2},\vec \kappa,\vec\Delta)$ 
is the skewed unintegrated gluon distribution,
with $x_1 \not = x_2$ being the fractions of the proton's momentum
carried by the uppermost gluons.
The appearance of skewed (or generalized) parton distributions
is characteristic for scattering processes that change the mass/virtuality
of the projectile \cite{GPD,GPDcollinear}.
In the $k_t$-factorization approach, the skewness is transferred to
the unintegrated distributions. The real part of the amplitude
can be reconstructed from the imaginary part (\ref{ampl}) using analyticity.

The twist analysis of helicity amplitudes (\ref{ampl}) was performed
in \cite{in99}. The principal findings were: (1) the longitudinal cross section
was found to be particularly suppressed in comparison with $S$-wave
meson production, which translated into abnormally small $\sigma_L/\sigma_T$ ratio,
and (2) the $s$-channel helicity non-conserving (SCHNC) amplitudes played
much more important role than in the case of $\rho$.
A comparison of $V_D$ and the spin-3 ground state of the same quarkonium $V_3$
performed in \cite{spin3} offered an insight into smallness of $\sigma_L(V_D)$
in terms of Clebsch-Gordan coefficients.

\section{Numerical results}

Here we present numerical results for the most interesting case of $\rho_D$,
which in this section we identify with the physical state $\rho(1700)$.

In order to integrate (\ref{ampl}) numerically, one needs
to specify models for the off-forward unintegrated gluon density and
for the vector meson wave function. The former was related
to the forward unintegrated gluon density, see \cite{review},
whose parametrizations were borrowed from \cite{in2000}.
The choice of a particular parametrization
affects the numerical results only marginally. 
For the vector meson wave function we used the appropriately normalized 
Gaussian Ansatz. The particular choice of the radial
wave function changes sizably only the overall normalization of the predicted
cross sections, but not their shapes.   
The experimental value of the leptonic decay width $\Gamma(\rho(1700) \to e^+e^-)$,
which was used to fix the size parameter of the wave function, 
is known very poorly, which significantly
affects the accuracy of our predictions. Below, we show 
numerical results using $\Gamma(\rho(1700) \to e^+e^-) = 0.14$--0.7 keV.
Below, we also show for comparison the cross sections of $\rho$ and $\rho_{2S}$
production. In both cases, we varied the corresponding leptonic decay widths
$\Gamma = (1\div 3)\Gamma_{\mbox{\tiny PDG}}$ to roughly control the uncertainty
of numerical results.

\begin{figure}[!htb]
   \centering
\includegraphics[width=12cm]{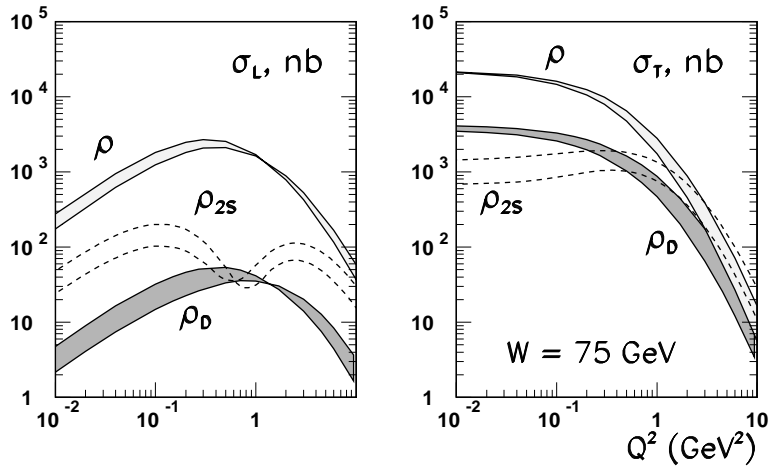}
\caption{The $Q^2$-dependence of the longitudinal (left plot) and transverse (right plot) $\rho_D$ 
production cross sections (dark shade regions). For comparison, we also show our results 
for the $\rho$ (light shaded regions) and $\rho_{2S}$ (dashed curves). 
The shaded regions show the sensitivity of the results
to the variation of the leptonic decay width.}
   \label{fig-q2dep}
\end{figure}

\begin{figure}[!htb]
   \centering
\includegraphics[width=8cm]{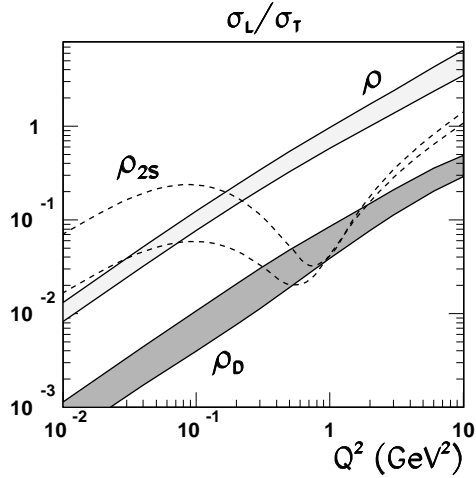}
\caption{The $Q^2$-dependence of $R = \sigma_L/\sigma_T$ ratio for $\rho_D$, $\rho$ and $\rho_{2S}$.
Notation is the same as in Fig.~\ref{fig-q2dep}.}
   \label{fig-rlt}
\end{figure}

\begin{figure}[!htb]
   \centering
\includegraphics[width=12cm]{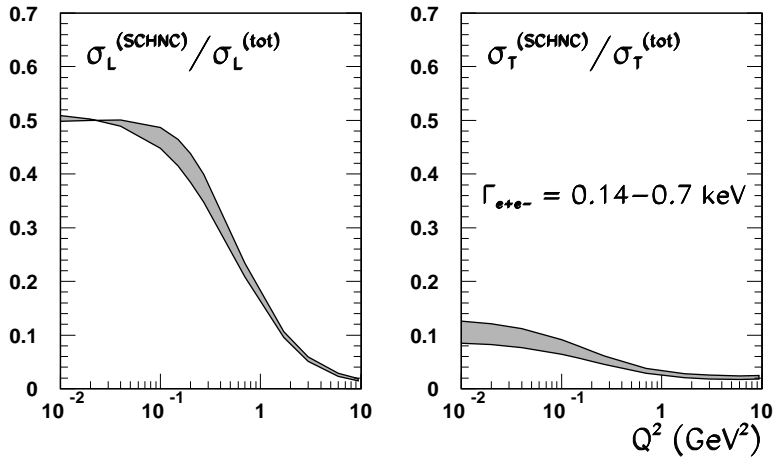}
\caption{The relative weight of the $s$-channel helicity non-conserving contributions
to longitudinal (left plot) and transverse (right plot) $\rho_D$ production 
cross sections as functions of $Q^2$.
The shaded regions correspond to $\Gamma(\rho(1700) \to e^+e^-) = 0.14$--0.7 keV.}
   \label{fig-schnc}
\end{figure}

In Fig.~\ref{fig-q2dep} we show the longitudinal 
and transverse cross sections for $\rho_D$ as well as for $\rho$ 
and $\rho_{2S}$. At the photoproduction point,  
$\sigma_T(\rho_D)/ \sigma_T(\rho)\approx 0.2$, and it
increases slightly towards higher $Q^2$. 
The cross section of $\rho_D$ production by longitudinal photons is always small,
$\sigma_L(\rho_D)/\sigma_L(\rho) \sim 0.01$--0.03, in agreement
with qualitative studies \cite{in99}. 
This translates into anomalously small longitudinal-to-transverse ratio 
$R(\rho_D) \equiv \sigma_L(\rho_D)/\sigma_T(\rho_D) \sim 0.1 R(\rho)$,
which is illustrated by Fig.~\ref{fig-rlt}.

Note that the $\rho_{2S}$ production cross sections show very different patterns
of $Q^2$ behavior, see Fig.~\ref{fig-q2dep} and Fig.~\ref{fig-rlt}. 
Comparison of $\sigma(\rho_{2S})$ and $\sigma(\rho_D)$
shows that at small $Q^2$, and in particular, at the photoproduction point,
$\rho_D$ should dominate over $\rho_{2S}$. 
At moderate $Q^2 \sim 1$ GeV$^2$ both mesons are expected to be produced
at comparable rates, and at $Q^2 \gsim 5$ GeV$^2$ 
the $\rho_{2S}$ production should take over. 

Fig.~\ref{fig-schnc} demonstrates the role of SCHNC transitions in $\rho_D$
production. In accordance with expectations, 
at small $Q^2$ the $s$-channel helicity violating amplitudes
generate a sizable portion of both longitudinal and transverse cross sections.
As $Q^2$ grows, the SCHNC effects gradually die out,
however, their magnitude is still larger than SCHNC observed 
in production of ground state $\rho$ mesons \cite{SCHNCHERA}.
At small $Q^2$, the presence of strong helicity violating amplitudes noticeably modifies
the $t$-dependence of the differential cross sections at $|t|\gsim 0.3$ GeV$^2$.


We also checked the energy dependence of the $\rho_D$ 
production cross section, which can be conveniently parametrized 
with simple power law, $\sigma(W)\propto W^\delta$. 
We checked the $Q^2$ behavior of the exponent $\delta$
and found it somewhat smaller 
than the corresponding exponent for $\rho$ production.
This can be traced back to the larger color dipole sizes probed in $\rho_D$
production and to the enhanced contribution from larger $|t|$ region,
although the effect is not as dramatic as in the case of spin-3 meson production, 
see \cite{spin3}.

\section{Testing purity of the $D$-wave}

If the $\rho(1450)$ and $\rho(1700)$ were well separated eigenstates of the $q\bar q$
angular momentum, then using $Q^2$ behavior of their diffractive production cross sections, 
Fig.~\ref{fig-q2dep}, one could easily tell radial from orbital excitation.
In reality one has to account for a possibly strong mixing between
the radial and orbital excitations as well as for their significant overlapping. 
The legitimate question is whether diffractive production can be used
under these circumstances to probe the structure of $\rho(1450)$ and $\rho(1700)$.
Analysis of this type was conducted in \cite{dosch}, where 
authors considered both $\rho(1450)$ and $\rho(1700)$
as states composed of a pure $2S$ and of anything else
(orbital excitation, hybrid etc.). They determined the mixing angle
to be $\theta = 41.2^\circ$, and were able to describe the scarce experimental
data available. With too many assumptions 
($\sigma(\rho_D) = 0$, strict SCHC etc.) and poorly known parameters,
the precision of the results obtained in \cite{dosch} seems to us doubtful. 

We agree with \cite{dosch} that one should study not $\rho(1450)$ and $\rho(1700)$
separately but $\pi^+\pi^-$ or $4\pi$ final state in the entire region
of multipion invariant mass $M = 1.2$--$1.8$ GeV.
In our opinion, two quantities are particularly useful for determination of
the internal structure of $\rho(1450)$ and $\rho(1700)$:
(i) the $Q^2$ variations of the invariant mass shape, and (ii) the level of 
$s$-channel helicity violation. Both quantities should be studied 
in the region of small-to-moderate $Q^2$, precisely where the $\rho(2S)$ cross section
is expected to have strong variations due to the node effect.
 
\begin{figure}[!htb]
   \centering
\includegraphics[width=12cm]{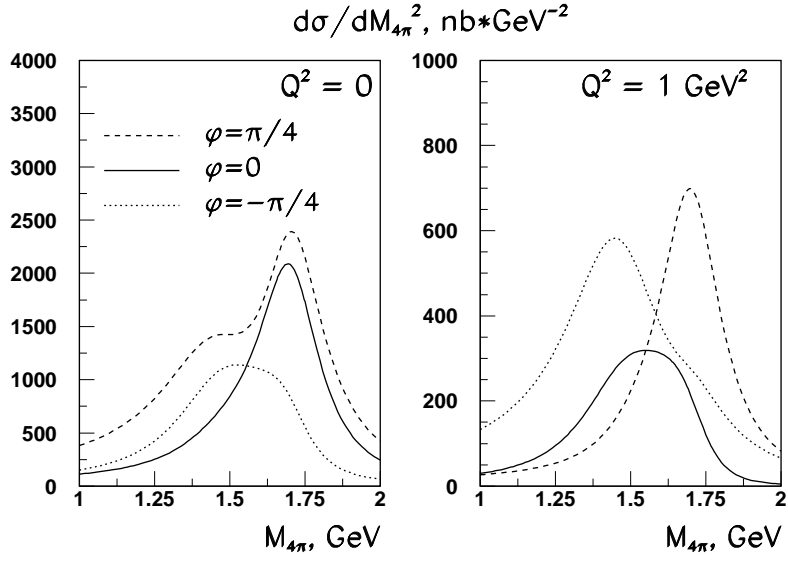}
\caption{Typical $M_{4\pi}$ dependence of diffractive $4\pi$ production cross section
in case of several values of mixing angle and for two values of photon virtuality:
$Q^2=0$ (left plot) and  $Q^2=1$ GeV$^2$ (right plot).}
   \label{fig-mix}
\end{figure}

\begin{figure}[!htb]
   \centering
\includegraphics[width=12cm]{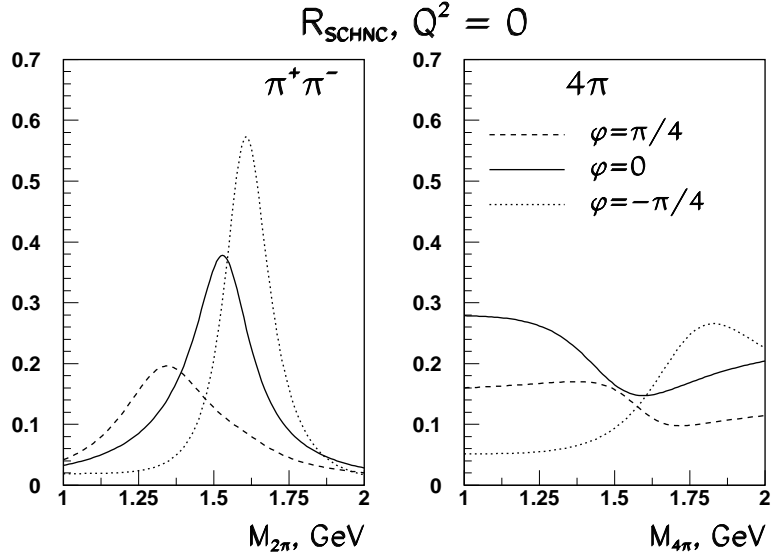}
\caption{The fractions of SCHNC contribution $\RSCHNC$ to the diffractive $\pi^+\pi^-$ (left plot) 
and $4\pi$ (right plot) photoproduction as functions of the final state invariant mass.
On each plot, three different curves shown correspond to no $2S/D$ mixing (solid lines),
mixing with $\phi=\pi/4$ (dashed lines) and mixing with $\phi=-\pi/4$ (dotted lines).}
   \label{fig-mix-schnc}
\end{figure}

We consider simple $2S$/$D$-wave mixing in the $\rho$ system and represent 
the two $\rho'$ states as
\be
|\rho(1450) \rangle = \cos\phi |2S\rangle + \sin\phi |D\rangle\,,\quad
|\rho(1700) \rangle = -\sin\phi |2S\rangle + \cos\phi |D\rangle\,.\label{mixing}
\ee
The total amplitude for a give final state $f = 2\pi,\, 4\pi$ with invariant mass $M_f$ 
is written as
\be
A = A_\rho D_\rho(M_f) + A_{\rho(1450)} D_{\rho(1450)}(M_f) + A_{\rho(1700)} D_{\rho(1700)}(M_f)\,,
\ee
where $A_i$ are the production amplitudes of each resonance calculated within the same 
approach and $D_i$ are the corresponding Breit-Wigner factors.
In Fig.~\ref{fig-mix} we show typical invariant mass spectra of the $4\pi$ states
for $Q^2 = 0$ and $Q^2 = 1$ GeV$^2$. The solid curves correspond to no mixing,
while dashed and dotted curves correspond to $\phi = \pi/4$ and $\phi = -\pi/4$.
The $M_{4\pi}$-shape at fixed $Q^2$ and its variation with $Q^2$ growth 
show interesting dependence on the mixing angle. 

Another way to separate radial/orbital excitations is to study the magnitude of the
$s$-channel helicity non-conservation 
$\RSCHNC = 1 - \sigma(\mbox{SCHC})/\sigma(\mbox{full})$
as a function of the final state invariant mass $M_f$.
Although at the photoproduction point both $\rho_{2S}$ and $\rho_D$ have
roughly $\sim 10$--$15\%$ of their cross sections coming from SCHNC amplitudes,
their interference pattern is sensitive to the mixing.
Fig.~\ref{fig-mix-schnc} shows values of $\RSCHNC$ in the cases of $\pi^+\pi^-$
(left plot) and $4\pi$ (right plot) final states in the photoproduction limit.
For these plots, we took $Br(\rho(1450)\to \pi^+\pi^-) = Br(\rho(1700)\to \pi^+\pi^-) = 0.2$.
In the $\pi^+\pi^-$ case, the helicity violating amplitudes come from 
excited states, while the helicity conserving amplitudes have a non-trivial
interference between $\rho$ and excited states. This produces a peak in $\RSCHNC(M_{2\pi})$
plot, whose position is sensitive to the mixing angle.
In the $4\pi$ final state, there is no $\rho$ contribution, and
the SCHNC is substantial throughout the whole $M_{4\pi}$ region shown,
its interference pattern again sensitive to the mixing. 
We underline that Figs.~\ref{fig-mix},~\ref{fig-mix-schnc} show only typical patterns
to look for. Exact predictions are difficult due to too many poorly known 
parameters in the game.

\section{Comparison with experimental data and future possibilities}

In 1980's the Omega Collaboration has studied diffractive photoproduction
of $\rho'(1600)$ meson in $\pi^+\pi^-$ and $4\pi$ channels.
From the published results $\sigma(\rho'\to\pi^+\pi^-)/\sigma(\rho) = 0.01 \pm 0.002$
\cite{aston1980} and $\sigma(\rho'\to 4\pi) = 0.7 \pm 0.2$ $\mu$b \cite{atkinson1981} 
one can estimate the $\rho'(1600)$ photoproduction cross section as roughly 1 $\mu$b.
Although these results were reanalyzed in \cite{tworhoprimes} 
in terms of two excited states $\rho(1450)$ and $\rho(1700)$,
we find it more secure to compare our results with the original data.

As shown on Fig.~\ref{fig-q2dep}, our predictions for the photoproduction cross sections
are $\sigma(\rho_{2S}) \sim 4$ $\mu$b, $\sigma(\rho_D) \sim 1$ $\mu$b.
However, since $\rho(1450)$ and $\rho(1700)$ are broad interfering peaks,
the integration of $d\sigma/dM_{4\pi}$ within the region $M_{4\pi} = 1.2$--1.8 GeV
yields $\sigma(\rho(1450)+\rho(1700)) \sim 1.5$--3 $\mu$b, depending on the 
$2S/D$ mixing angle, which is not far from experiment.

As discussed above, detailed studies of the $2\pi$ or $4\pi$ mass spectrum,
in particular, the level of $s$-channel helicity violation,
can help resolve the $S$/$D$-wave structure of the $\rho(1450)$ and $\rho(1700)$.
Extraction of SCHNC amplitudes relies on reconstruction of the angular
distribution of the final multipion state. For the $\pi^+\pi^-$ state this
procedure was described in \cite{SW}, while for the $4\pi$ state 
more analysis in the spirit of \cite{atkinson1981} should be done.
Note that the conclusion made in the latter paper that at least half 
of all $4\pi$ events under the peak come from an SCHC mechanism
does not contradict the results of our calculations.

When extracting $\rho'$ from the multipion final states,
one must subtract from the data the $\rho_3(1690)$
contribution, which is not negligible and which is predicted to have extremely 
large SCHNC contibutions, see \cite{spin3}.
Such a separation would be possible if one either
performs the full partial wave analysis, which is more suitable 
for the $\pi^+\pi^-$ final state, 
or extracts a specific signal in the multipion final states, 
like $\rho_3 \to a_2(1320)\pi\to \eta\pi^+\pi^-$.

We also briefly comment on the recent observation by Focus Experiment \cite{focus}
of a very statistically significant enhancement in the diffractive photoproduction 
of $K^+K^-$ pairs at $M_{KK} = 1750$ MeV.
The experiment ruled out the possibility to explain this enhancement
with $\phi(1680)$ meson, which is believed to be mainly $2S$ excitation
of $\phi$. One could speculate if this new resonance could be 
the orbitally excitated $\phi$ meson. 
If so, our calculations predict the photoproduction cross section 
$\sigma(\phi_D) \sim 200$--300 nb, which is $\sim 1/5$ of the $\phi$ photoproduction
cross section calculated in the same model.
Unfortunately, the experimental paper does not give the measurement results
for the cross section.

\section{Conclusions}

In this paper we considered diffractive production of orbitally 
excited vector mesons in DIS, with a major focus on the $D$-wave $\rho'$ state.
The calculations were performed within the $k_t$-factorization framework.
Although the predictions for the absolute values of the cross sections
are plagued by large uncertainties of the input parameters, we find that several
qualitative conclusions are stable:
\begin{itemize}
\item The $\rho_D$ production cross section
is roughly an order of magnitude smaller than the $\rho$ production cross section.
At small $Q^2$, $\sigma(\rho_D)$ is larger than $\sigma(\rho_{2S})$, casting
doubt on the relevance of $\sigma(\rho_D)=0$ assumption used in \cite{dosch}.
\item Studying $\sigma_L$ and $\sigma_T$ separately, we observe strong 
suppression of the longitudinal cross section, which translates into very small 
ratio of $R = \sigma_L/\sigma_T$, as was anticipated in \cite{in99}. 
\item  The role of the $s$-channel helicity violation is more important in $D$-wave
state production than in the ground state production, especially at small $Q^2$.
\end{itemize}
Besides, our results confirm general expectations that the experimental study 
of diffractive multipion production in the invariant mass region 
of $\rho(1450)/\rho(1700)$ interference can be a key to the 
$2S$-wave/$D$-wave assignment to these two mesons. Study of the $Q^2$ variation
of $M_{4\pi}$ shapes of the cross section as well as checking the level 
of $s$-channel helicity violation should be of much help.\\

The work of I.P.I. is supported by the INFN Fellowship, and partly by INTAS and grants 
RFBR 05-02-16211 and NSh-2339.2003.2.

\end{document}